\documentclass[]{spie}  

\usepackage{amsmath,amsfonts,amssymb}
\usepackage{graphicx}
\usepackage[colorlinks=true, allcolors=blue]{hyperref}

\title{More data than you want, less data than you need: machine~learning approaches to starlight subtraction with MagAO-X}

\author[a]{Joseph D. Long}
\author[b]{Jared R. Males}
\author[b]{Laird M. Close}
\author[b,c,d,e]{Olivier Guyon}
\author[b,f]{Sebastiaan Y. Haffert}
\author[g]{Alycia J. Weinberger}
\author[c]{Jay Kueny}
\author[b]{Kyle Van Gorkom}
\author[c]{Eden McEwen}
\author[b]{Logan Pearce}
\author[c]{Maggie Kautz}
\author[b]{Jialin Li}
\author[c]{Jennifer Lumbres}
\author[h]{Alexander Hedglen}
\author[i]{Lauren Schatz}
\author[j]{Avalon McLeod}
\author[k]{Isabella Doty}
\author[b]{Warren B. Foster}
\author[b]{Roswell Roberts}
\author[b]{Katie Twitchell}

\affil[a]{Center for Computational Astrophysics, Flatiron Institute, New York, New York, USA}
\affil[b]{Steward Observatory, The University of Arizona, Tucson, Arizona, USA}
\affil[c]{James C. Wyant College of Optical Sciences, University of Arizona, Tucson, Arizona, USA}
\affil[d]{Subaru Telescope, National Astronomical Observatory of Japan, Hilo, Hawaii, USA}
\affil[e]{Astrobiology Center, National Institutes of Natural Sciences, Tokyo, Japan}
\affil[f]{Leiden Observatory, Leiden University, Netherlands}
\affil[g]{Earth and Planets Laboratory, Carnegie Institution for Science, Washington D.C., USA}
\affil[h]{Northrop Grumman Corporation, Rolling Meadows, Illinois, USA}
\affil[i]{Starfire Optical Range, Kirtland Air Force Base, Albuquerque, New Mexico, USA}
\affil[j]{Draper Laboratory, Cambridge, Massachusetts, USA}
\affil[k]{University of Rochester, Rochester, New York, USA}

\authorinfo{Corresponding author: Joseph D. Long (\href{mailto:jlong@flatironinstitute.org}{jlong@flatironinstitute.org})}

\begin{document} 
\maketitle

\begin{abstract}
High-contrast imaging data analysis depends on removing residual starlight from the host star to reveal planets and disks.  Most observers do this with principal components analysis (i.e. KLIP) using modes computed from the science images themselves. These modes may not be orthogonal to planet and disk signals, leading to over-subtraction. The wavefront sensor data recorded during the observation provide an independent signal with which to predict the instrument point-spread function (PSF). MagAO-X is an extreme adaptive optics (ExAO) system for the 6.5-meter Magellan Clay telescope and a technology pathfinder for ExAO with GMagAO-X on the upcoming Giant Magellan Telescope. MagAO-X is designed to save all sensor information, including kHz-speed wavefront measurements. Our software and compressed data formats were designed to record the millions of training samples required for machine learning with high throughput. The large volume of image and sensor data lets us learn a PSF model incorporating all the information available. This allows us to probe smaller star-planet separations at greater sensitivities, which will be needed for rocky planet imaging.
\end{abstract}


\section{INTRODUCTION}
\label{sec:intro}

Of all the confirmed exoplanets discovered to date, only a handful can be directly imaged\cite{nasaexoplanetscienceinstitutePlanetarySystemsTable2020}. Exoplanet and disk direct imaging, and more broadly direct characterization (e.g. by spectroscopic or polarimetric measurents) allows us to study the light emitted or reflected by the object itself. However, direct imaging of exoplanets remains exceptionally difficult with current-generation instruments and telescopes. The Astro2020 decadal survey\cite{astro2020} identified improving our capabilities in exoplanet direct imaging as a strategic priority.

Both ground-based and space-based astronomical high-contrast imaging observations are impeded by ``speckles'', the irregular diffractive structure of the point-spread function (PSF) from wavefront errors in the imaging system that cause its response to deviate from that of an image of an ideal flat wavefront. Accurate characterization of these speckles is necessary to subtract residual starlight (or to jointly model the residual starlight and a companion or extended structure). Unfortunately, the speckle structure is not static, even on fairly short timescales\cite{malesMysteriousLivesSpeckles2021}. Speckles may appear and disappear over the course of an observation, and they limit our ability to detect faint companions. Thanks to this temporally- and spatially-correlated noise structure, the expected signal-to-noise ratio does not improve with time directly but rather with the number of realizations of the speckle pattern (which has a characteristic timescale over which it evolves). As a result, accurately characterizing these speckles with post-processing algorithms is an active area of research\cite{cantalloubeExoplanetImagingData2020}.

The characteristic angular scale of a PSF and a speckle alike is $\lambda/D$ where $\lambda$ is the wavelength and $D$ the telescope primary aperture diameter. Future telescopes will increase $D$, shrinking the speckle-limited region around the stellar PSF core. The transition from thermal infrared to visible light adaptive optics is happening today, shrinking the wavelengths used for high-contrast imaging from multiple microns to hundreds of nanometers. The MagAO-X extreme adaptive optics system is at the forefront of this shift, providing high-Strehl images in visible wavelengths for exoplanet science, while also serving as a technology pathfinder for future extremely large telescopes (ELTs) like the Giant Magellan Telescope (GMT).

To maximize scientific productivity with MagAO-X, we want to detect companions and disk structure within the speckle-limited region closest to the star. However, the evolving speckle pattern presents a well-known challenge to starlight subtraction. Differential imaging techniques are used to decorrelate companion and disk signals from residual starlight. Angular differential imaging (ADI)\cite{maroisAngularDifferentialImaging2006} exploits the apparent rotation of the field of view as the target crosses the sky, in which companion and disk light will rotate but the shape of the PSF will remain fixed. Spectral differential imaging (SDI) uses wavelength diversity\cite{sparksImagingSpectroscopyExtrasolar2002,biller_sdi} to separate residual starlight from the signals of interest. Polarization differential imaging (PDI) uses polarization as another source of diversity, as scattered light from a disk will have a polarized component that the host star's light lacks\cite{perrinPolarimetryGeminiPlanet2015,schmidSPHEREZIMPOLHigh2018}.

Data-driven methods to estimate the pattern from the science images themselves are prone to over-subtracting the signal of interest as well as the starlight\cite{milliImpactAngularDifferential2012,pueyoDetectionCharacterizationExoplanets2016}. The MagAO-X instrument runs at 2--3.6 kHz\cite{2018SPIE10703E..09M,2022SPIE12185E..09M,males2024}, and can store all the wavefront sensor (WFS) images taken in an hours-long observation for later analysis. We use these images as the inputs to an interpretable machine learning model that produces optical path difference and pupil amplitude maps. From these maps, a differentiable physical optics model constructs an electric field for a notional pupil and propagates to the focal plane and a simple detector model.

We report the progress to date in using saved MagAO-X wavefront sensor telemetry and science data to learn a transformation that captures the ``unknown unknowns'' in the AO imaging system for improved speckle modeling and, eventually, improved exoplanet detection limits.

\section{OVERVIEW}

The de-facto starlight subtraction algorithm in high-contrast imaging remains the KLIP/PCA\cite{klip,amaraPYNPOINTImageProcessing2012} method to empirically compute a basis of eigen-images, exploiting a low-rank representation of the space of speckle patterns to exclude the astrophysical signals of interest (i.e. exoplanets and disks) from the PSF model. Among the shortcomings of this approach are poor asymptotic scaling in the amount of data ($O(N^4)$ in the number of frames, $N$, see Ref.~\citenum{longUnlockingStarlightSubtraction2021a}) as well as a tendency to subtract more of the signal of interest as the rank of the subspace increases\cite{pueyoDetectionCharacterizationExoplanets2016}.

The broad utility and success of the empirical KLIP/PCA approach to modeling residual starlight, when compared to a physical forward modeling approach, indicates there are more ``unknown unknowns'' than we can hope to capture with any simple physical model. While space-based observatories benefit from relative stability and can be modeled in detail, ground-based adaptive optics systems must contend with Earth's changing atmosphere as the first optic in the imaging system. Extreme adaptive optics systems for exoplanet direct imaging and characterization run at multiple kiloHertz\cite{guyonExtremeAdaptiveOptics2018}, compensating for atmospheric aberrations in closed-loop by sensing and correcting the incoming wavefront. The photons hitting the wavefront sensor are, to a good approximation, entirely stellar in origin. (In any case, a photon arriving from a closely-separated companion will experience almost identical turbulence and will not contaminate the wavefront sensing signal.) The wavefront sensor frames thus give us information about the state of the PSF---and its evolution within a science frame---that is entirely independent of a potential companion or disk signal.

We put this property to use in the development of a machine learning model to estimate a post-facto reconstruction of the wavefront evolution during a science exposure. These wavefronts form the input to a differentiable physical optics model, which propagates by a matrix Fourier transform\cite{soummerFastComputationLyotstyle2007} to an over-sampled detector focal plane. The resulting image is binned to approximate finite pixel size, and a simple convolution applies a learned kernel to the images to capture detector inter-pixel effects. The resulting science image is then compared with the ground truth science image to compute the loss function, the scalar-valued function which is automatically differentiated by the JAX algorithmic differentiation library\cite{jax2018github,deepmind2020jax}. By performing gradient descent with the Equinox\cite{kidger2021equinox} machine learning framework and Optax optimizer library\cite{deepmind2020jax}, we learn the neural network weights and biases that transform wavefront sensor images into optical path difference (OPD) and amplitude errors. The neural network learns the nonlinear transformations into a latent space and then back to form coefficients of a modal basis set, which is then combined to get the OPD and amplitude maps.

\section{MODEL}

\begin{figure}[ht]
\begin{center}
\includegraphics[height=5cm]{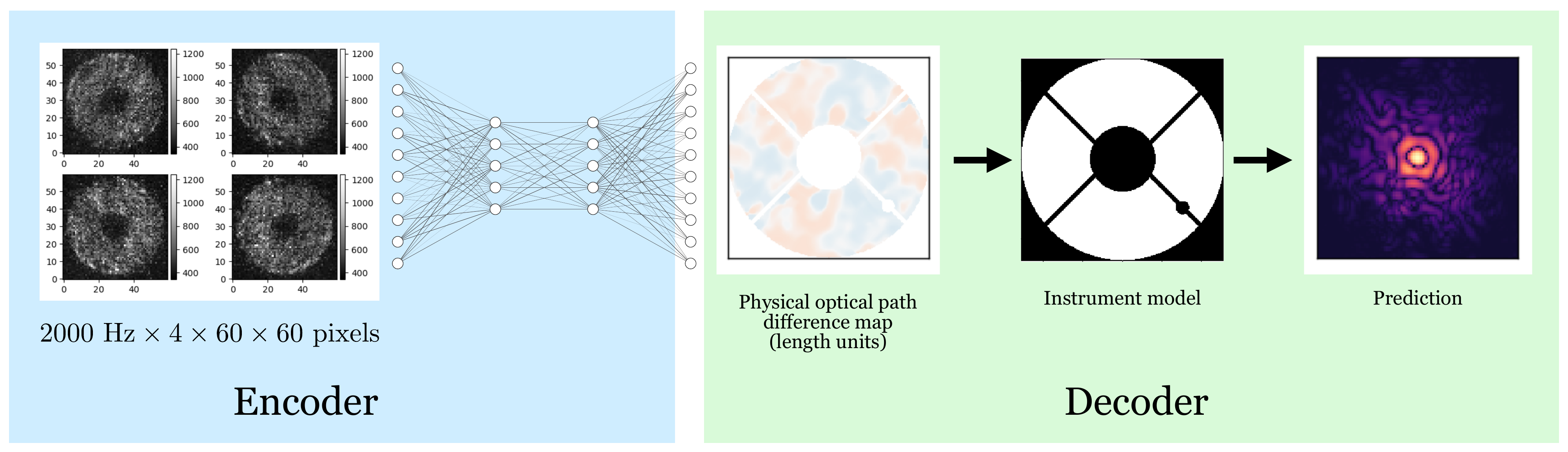}
\end{center}
\caption[Encoder-decoder architecture]{ \label{fig:encoder-decoder} A simplified diagram of the model architecture. The ``encoder'' stage uses a learned transformation to convert the input wavefront sensor frames into a low rank representation. The vector produced at the end of the encoder stage provides coefficients to construct an OPD map from a modal basis set. The resulting OPD map and pupil mask can then be used to simulate a detector image. This transformation from low-dimensional back to high-dimensional is sometimes called ``decoding'' the representation the neural network has learned.}
\end{figure} 

The model takes some number of raw wavefront sensor frames and produces a single science focal plane image. It accomplishes this by applying a learned nonlinear transformation from wavefront sensor image space to a vector of modal basis coefficients, combining the modal basis according to these coefficients to produce an OPD map and a map of amplitude aberrations, and then calculating a PSF for multiple wavelengths on the correct sampling for the science focal plane. These are summed incoherently across wavelength and wavefront sensor frame index before being multiplied by an amplitude scale factor to match the detector count values. A simple single-parameter bias level model is added, and the result is compared to the focal plane image. A simplified diagram is shown in Figure~\ref{fig:encoder-decoder}.

\subsection{Data loading and pre-processing}

We load an observation sequence as a collection of detector time-series data cubes paired with time stamps. A single science camera frame will always correspond to multiple wavefront sensor frames, we group the wavefront sensor frames that were acquired during a single science frame. This is a metadata-only operation, so we do not need to load the entire wavefront sensing dataset from disk or network to do it. The resulting mapping of science frame indices to wavefront sensor frame indices is a ``data association'' in our code, and exposes an interface to load a single ``batch'' (some number of science frames and their associated WFS frames) from storage, optionally cropping the science frames to a smaller subframe around the PSF for memory efficiency and speed.

\subsection{Wavefront sensor to OPD and amplitude maps}

Drawing some inspiration from Landman \& Haffert\cite{landmanMakingUnmodulatedPyramid2024}, we begin by rearranging the four pupils of our pyramid wavefront sensor into four channels (i.e. $120 \times 120 \rightarrow 4 \times 60 \times 60$). A convolutional neural network layer is applied, with the weights of the four convolution kernels learned in the training process. The output, a single $60 \times 60$ channel, is passed through a rectified linear unit (ReLU) activation function and unraveled into a single vector.

\begin{figure}[ht]
\begin{center}
\includegraphics[width=0.9\textwidth]{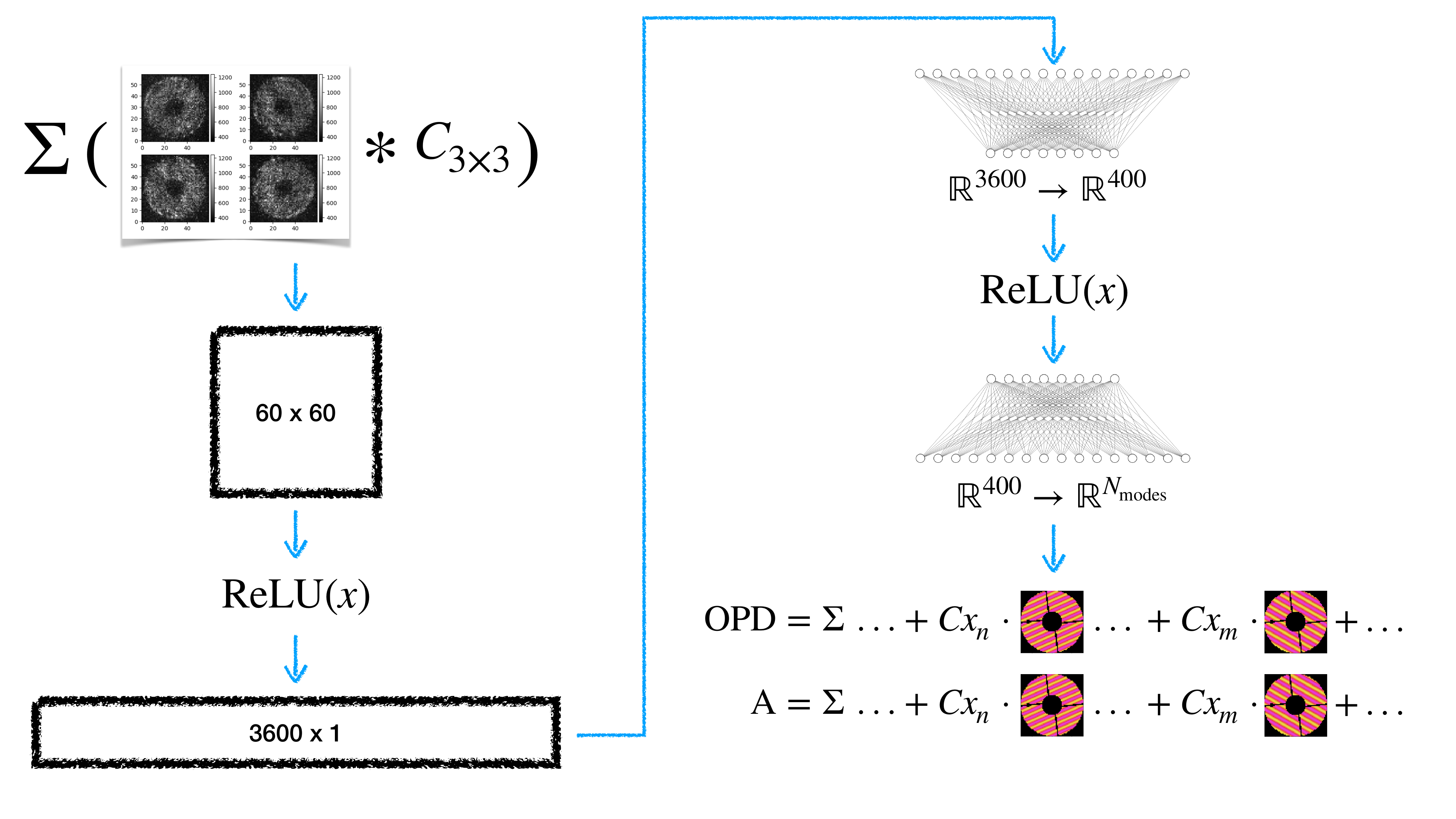}
\end{center}
\caption[Single WFS frame flowchart]{ \label{fig:single_wfs_flow} Process of converting a single wavefront sensor frame into an OPD map.}
\end{figure}

This vector in $\mathbb{R}^{3600}$ is passed through a fully-connected layer transforming it to a vector in $\mathbb{R}^{400}$ (a so-called ``bottleneck'') and then through a ReLU activation function. From this bottleneck, the next transformation is to a vector of modal coefficients. These modal coefficients are used to construct a map of OPD (to be turned into phase) and amplitude aberrations across the pupil. The process is illustrated in Figure~\ref{fig:single_wfs_flow}.

Both the bottleneck dimension and the number of modal coefficients are configurable parameters of the model, and we are continuing to experiment with both hyperparameter tuning and architecture improvements.

\subsection{Science focal plane}

\begin{figure}[ht]
\begin{center}
\includegraphics[width=0.9\textwidth]{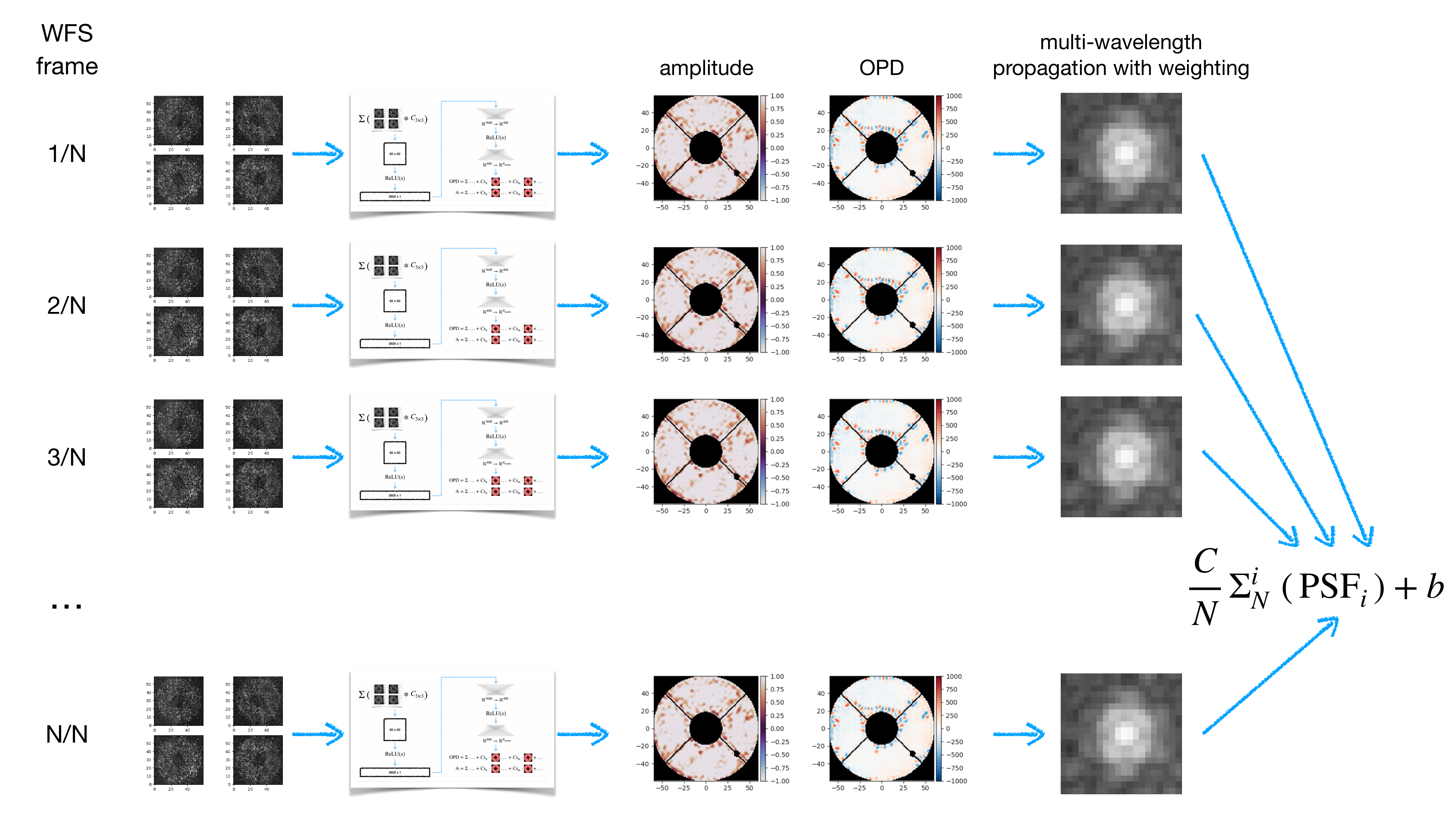}
\end{center}
\caption[Multiple WFS frames flowchart]{ \label{fig:all_wfs_flow} Overview of the process by which multiple WFS frames are combined to a single PSF. The $C$ variable represents the learned scale factor to match the counts in the detector images, and $b$ is the detector bias level in counts. Images are averaged to avoid linking the final flux scale factor to the number of WFS frames per science frame.}
\end{figure}

To generate a single science focal plane image requires processing the OPD maps corresponding to many wavefront sensor frames, simulating a PSF for each at multiple wavelength steps and incoherently adding them. By reordering the inner loop such that we iterate \emph{first} by wavelength step, \emph{then} by wavefront sensor frame index, we can reuse the ``plan'' matrices for that wavelength step and avoid a lot of unnecessary recomputation. The model learns a single amplitude scaling factor for each wavelength step, intended to capture variation in the source SED. The challenges with generalizing this to multiple targets will be explored in future work. The resulting PSFs are averaged rather than summed to avoid linking the final flux scale factor to the number of wavelengths propagated or the number of WFS frames per science frame. Finally, the summed PSF is rescaled to match the counts in the focal plane PSF (another target- and configuration-specific value to be generalized in future work) and a bias level is added. The process is illustrated in Figure~\ref{fig:all_wfs_flow}.

\subsection{Training}

We use JAX and the Optax optimizer library to learn the neural network weights and convolution kernels that produce the desired transformation. Non-convex optimization is as much an art as a science, and we have experimented with multiple optimizers, loss functions, and learning rates/schedules with many different combinations producing acceptable results. However, none has yet produced exceptional results, indicating we may need to revisit our architecture or incorporate additional components in our decoder model.

As representative examples, we have included some diagnostic plots from our training process. In this particular run, we used the Adam optimizer\cite{kingmaAdamMethodStochastic2015} with Nesterov momentum supplied by Optax to optimize a Huber loss function\cite{Huber1992} of the form
\begin{equation}
\label{eq:huber_loss}
\mathrm{Huber}(\theta, \mathbf{T}_i) = \begin{cases} 
          \frac{1}{2} |\mathbf{P}_i(\theta) - \mathbf{T}_i|^2 & |\mathbf{P}_i - \mathbf{T}_i| \leq \delta \\
          \frac{1}{2} \delta^2 + \delta (|\mathbf{P}_i(\theta) - \mathbf{T}_i| - \delta) & |\mathbf{P}_i - \mathbf{T}_i| > \delta
       \end{cases}
\end{equation}
where $\theta$ are the model parameters, $\mathbf{P}(\theta)$ is the prediction for some model parameters, $\mathbf{T}(\theta)$ is the truth or target image, and $i$ is an index over the . $\delta = 1$ is a constant defining the range in which the loss behaves as an $L_2$ norm of the errors. Outside of that range, the loss function behaves like an $L_1$ norm.

\begin{figure}[ht]
\begin{center}
\includegraphics[height=5cm]{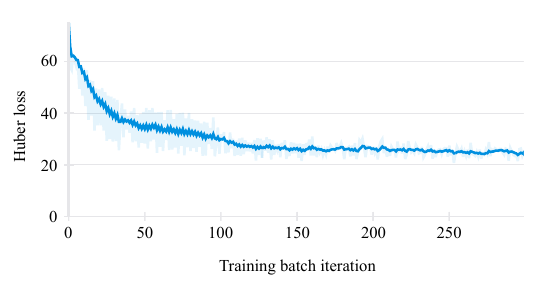}
\end{center}
\caption[Evolution of the loss function over the course of training]{ \label{fig:training-loss} Evolution of the loss function over the course of training. The dark blue line has been smoothed to better show the overall trend. The light blue line is the reported average Huber loss (Eq.~\ref{eq:loss}) at each iteration.}
\end{figure}

\begin{figure}[ht]
\begin{center}
\includegraphics[width=0.9\textwidth]{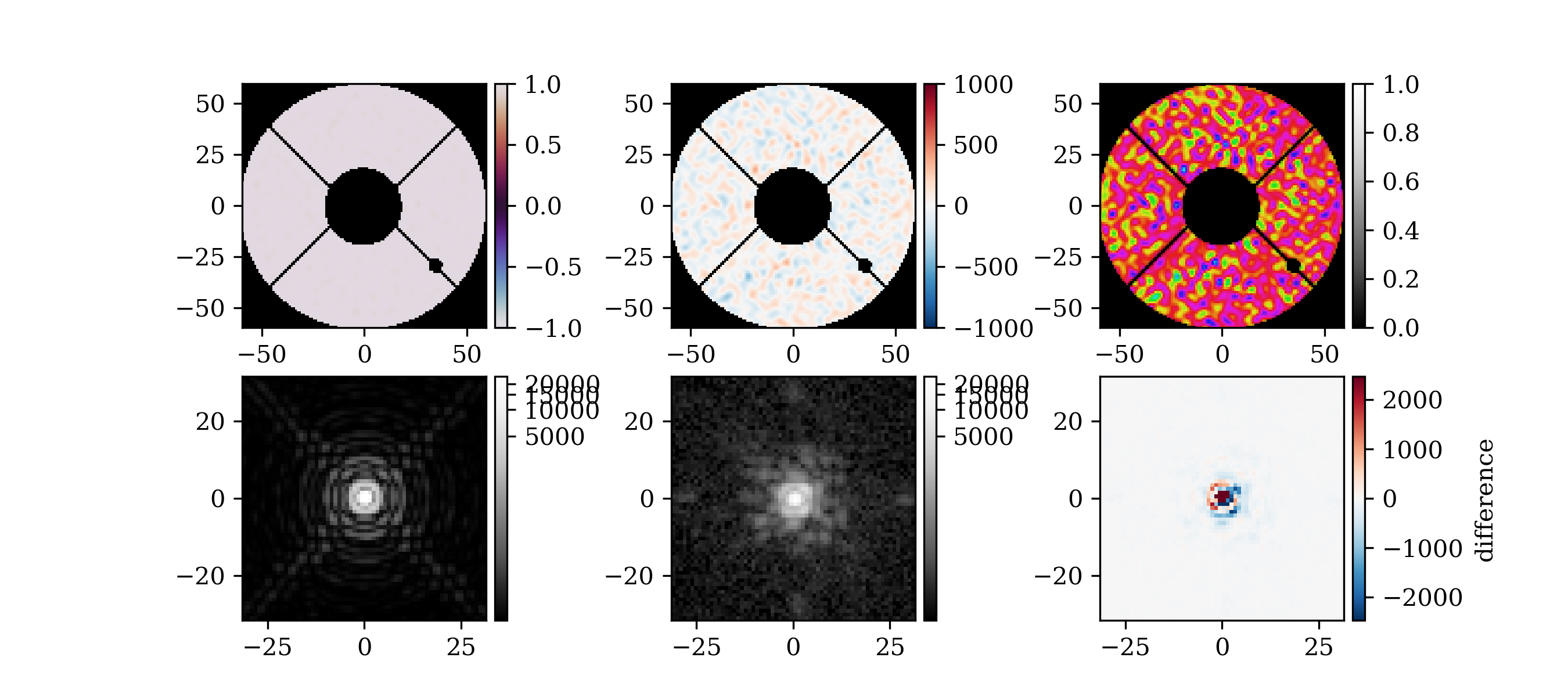}
\includegraphics[width=0.9\textwidth]{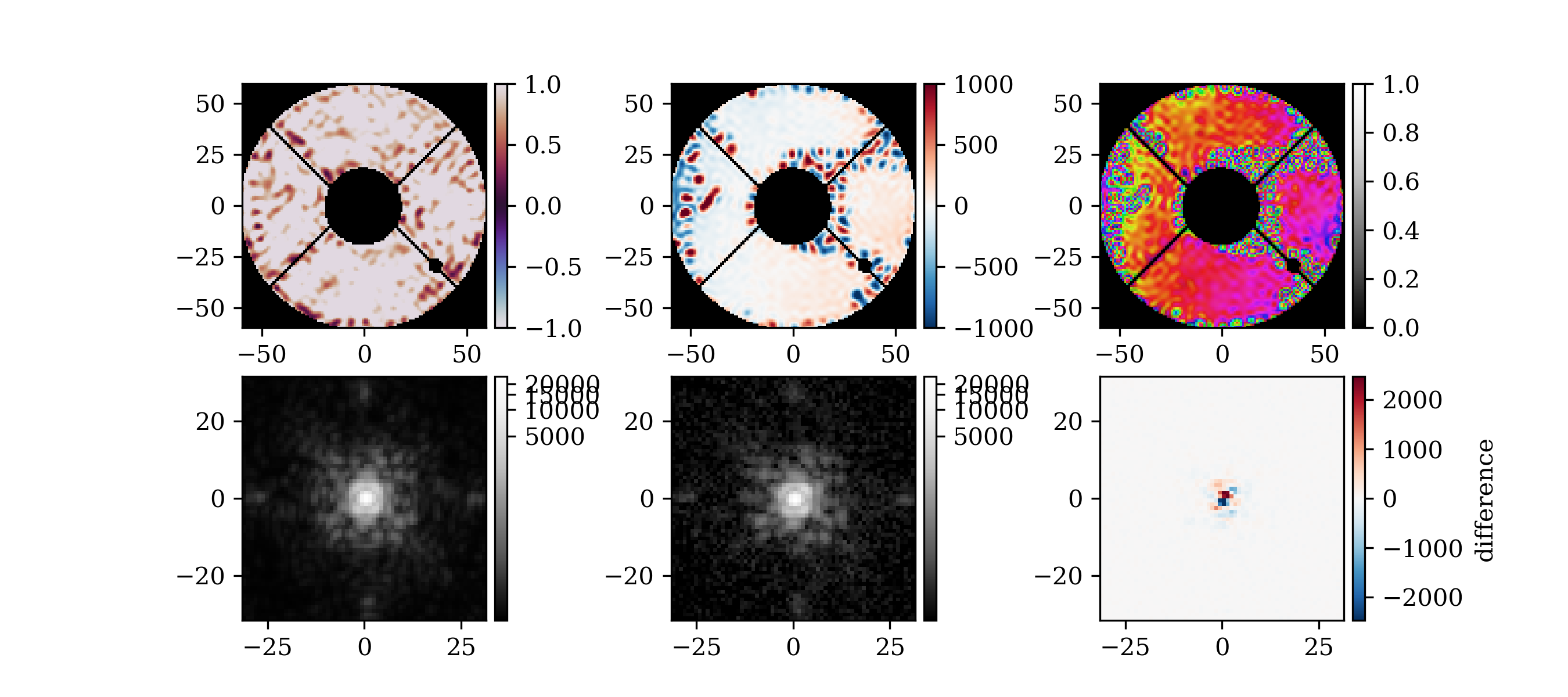}
\end{center}
\caption[Initial and final states of the model in the 300-iteration training run]{ \label{fig:initial-final} Initial and final states of the model in the 300-iteration training run shown in Figure~\ref{fig:training-loss}. The first six panels show (in left-to-right, top-to-bottom order): amplitude aberrations as fractional throughput in $[-1, 1]$, OPD in nanometers, the electric field color-coded by hue angle to represent phase, the model PSF from a single set of wavefront sensor frames corresponding to a single science frame (log-scaled), the target science frame (log-scaled), and a difference plot in units of counts. The second set of six panels shows the same after the parameters of the model have been optimized for 300 iterations. We do not believe the amplitude aberrations correspond to actual opaque material in the telescope pupil, but rather the model representing some phase errors at intermediate foci as a combination of phase and amplitude aberrations.}
\end{figure}

The function is applied element-wise across all the pixels in all the target images in the batch, so the actual scalar-valued loss function is the mean of the Huber loss across all the predicted pixels in a batch
\begin{equation}
    \label{eq:loss}
    \mathcal{L}(\theta,\mathbf{T}) = \frac{1}{N}\sum_{i=0}^N\ \mathrm{Huber}(\theta, \mathbf{T}_i).
\end{equation}
The value of the loss function across training batch iterations is shown in Figure~\ref{fig:training-loss}. The initial and final states of the model are illustrated by Figure~\ref{fig:initial-final}.

\section{CONCLUSION}

Algorithmic differentiation and accelerated linear algebra have been transformational in the development of machine learning models across disciplines. However, their workings are a largely a ``black box'' even to those who design them. By developing an accelerated, auto-differentiated physical optics model as part of the ``decoder'' stage of a deep learning model, we can interpret the intermediate representations learned by the model in physical terms and assess the plausibility of the model outputs.

The high number of wavefront sensor frames produced during a single science frame means the model must be developed with memory use in mind, and training batch sizes are kept small as a consequence. Certain information is simply not present in the wavefront sensor frames, notably any non-common-path aberration. The model must include enough flexibility to learn this, e.g., as a bias on the final layer of modal coefficients.

As this work is still in the early stages, we will continue to iterate on the model architecture to achieve better performance. By targeting a subset of the modes enabled on MagAO-X, we have a simplified version of the PSF reconstruction problem on which to assess the performance of our model. However, for scientific applications, we will need to support more realistic MagAO-X configurations. This will involve adding coronagraph support to the model, at a minimum. There is also low-hanging fruit to be picked with the simultaneous spectral differential imaging mode of MagAO-X. We expect the OPD map generated by the model to be equally valid for both science channels, which are typically configured with different filters. Applying the model simultaneously to the science channels in training will exploit this wavelength diversity to break degeneracies and improve the fidelity of the learned OPD and amplitude maps.

The MagAO-X instrument is particularly well-suited to this kind of post-processing, as it was designed to save every bit of telemetry data in operation. We have not even begun to investigate the possibilities of incorporating low-order wavefront sensing loop data or the  deformable mirror command stream into the PSF reconstruction process. Initial results have shown promise, and we will continue improving the model and eventually apply it to starlight subtraction for exoplanet and disk science.

\acknowledgments
 
JDL gratefully acknowledges the support of the Flatiron Software Research Fellowship at the Flatiron Institute, a division of the Simons Foundation. We are very grateful for support from the NSF MRI Award \#1625441~(MagAO-X). The MagAO-X~Phase~II upgrade program is made possible by the generous support of the Heising-Simons Foundation.

\bibliography{report} 
\bibliographystyle{spiebib} 

\end{document}